\documentclass[aps,prl,reprint,twocolumn,superscriptaddress,showpacs]{revtex4-1}
\usepackage{graphicx}
\usepackage{verbatim}
\usepackage{mathrsfs}
\usepackage{gensymb}
\usepackage{color}
\usepackage{colortbl}
\usepackage{ulem}

\usepackage{amsmath,amsfonts,amssymb}
\usepackage{graphicx}

\definecolor{mygray}{gray}{.9}
\definecolor{mypink}{rgb}{.99,.91,.95}
\definecolor{mycay}{rgb}{.3,.75,.93}
\newcommand{\bee}{\begin{equation}}
\newcommand{\ee}{\end{equation}}
\def\3{2.8in}    
\def\2{2.5in}
\def\4{3.0in}\def \beq {\begin{equation}}
\def \eeq {\end{equation}}
\newcommand{\upcite}[1]{\textsuperscript{\textsuperscript{\upcite{#1}}}}

\makeatletter

    \def\CT@@do@color{%
      \global\let\CT@do@color\relax
            \@tempdima\wd\z@
            \advance\@tempdima\@tempdimb
            \advance\@tempdima\@tempdimc
    \advance\@tempdimb\tabcolsep
    \advance\@tempdimc\tabcolsep
    \advance\@tempdima2\tabcolsep
            \kern-\@tempdimb
            \leaders\vrule
                    \hskip\@tempdima\@plus  1fill
            \kern-\@tempdimc
            \hskip-\wd\z@ \@plus -1fill }
    \makeatother

\begin{document}

\title{Topological chiral kagome lattice}

 \author{Jing-Yang You}
 \affiliation{Department of Physics, National University of Singapore, 2 Science Drive 3, Singapore 117551}
 
  \author{Xiaoting Zhou}
   \email{physxtzhou@gmail.com}
  \affiliation{Department of Physics, Northeastern University, Boston, MA, 02115, USA}
 
  \author{Tao Hou}
 \affiliation{Division of Physics and Applied Physics, School of Physical and Mathematical Sciences, Nanyang Technological University, Singapore 637371, Singapore}
 
  \author{Mohammad Yahyavi}
 \affiliation{Division of Physics and Applied Physics, School of Physical and Mathematical Sciences, Nanyang Technological University, Singapore 637371, Singapore}

  \author{Yuanjun Jin}
 \affiliation{Division of Physics and Applied Physics, School of Physical and Mathematical Sciences, Nanyang Technological University, Singapore 637371, Singapore}
 
  \author{ Yi-Chun Hung}
    \affiliation{Department of Physics, Northeastern University, Boston, MA, 02115, USA}
        \affiliation{Institute of Physics, Academia Sinica, Taipei 115229, Taiwan}

   \author{Bahadur Singh}
 \affiliation{Department of Condensed Matter Physics and Materials Science, Tata Institute of Fundamental Research, Mumbai, India}
 
\author{Chun Zhang}
\affiliation{Department of Physics, National University of Singapore, 2 Science Drive 3, Singapore 117551}

\author{Jia-Xin Yin}
\affiliation{Department of Physics, Southern University of Science and Technology, Shenzhen, Guangdong 518055, China}

  \author{Arun Bansil}
  \affiliation{Department of Physics, Northeastern University, Boston, MA, 02115, USA}

\author{Guoqing Chang}
\email{guoqing.chang@ntu.edu.sg}
\affiliation{Division of Physics and Applied Physics, School of Physical and Mathematical Sciences, Nanyang Technological University, Singapore 637371, Singapore}

\begin{abstract}
Chirality, a fundamental structural property of crystals, can induce many unique topological quantum phenomena. In kagome lattice, unconventional transports have been reported under tantalizing chiral charge order. Here, we show how by deforming the kagome lattice to obtain a three-dimensional (3D) chiral kagome lattice in which the key band features of the non-chiral 2D kagome lattice -- flat energy bands, van Hove singularities (VHSs), and degeneracies -- remain robust in both the $k_z=0$ and $\pi$ planes in momentum space. Given the handedness of our kagome lattice, degenerate momentum points possess quantized Chern numbers, ushering in the realization of Weyl fermions. Our 3D chiral kagome lattice surprisingly exhibits 1D behavior on its surface, where topological surface Fermi arc states connecting Weyl fermions are dispersive in one momentum direction and flat in the other direction. These 1D Fermi arcs open up unique possibilities for generating unconventional non-local transport phenomena at the interfaces of domains with different handedness, and the associated enhanced conductance as the separation of the leads on the surface is increased. Employing first-principles calculations, we investigate in-depth the electronic and phononic structures of representative materials within the ten space groups that can support topological chiral kagome lattices. Our study opens a new research direction that integrates the advantages of structural chirality with those of a kagome lattice and thus provides a new materials platform for exploring unique aspects of correlated topological physics in chiral lattices.
\end{abstract}
\maketitle


In the past decade, there has been a surge of research interest in topological materials characterized by non-trivial geometric phases~\cite{Hasan2010,Yu2010,Fu2007,Chang2013,Qi2011,Burkov2016,Bansil2016,Zhong2016,Bradlyn2017,Schindler2018,Wen2017,Armitage2018,Lv2021,Hasan2021}. Another parallel area of focus has revolved around correlated systems with strong interactions~\cite{Lee2006,Zhou2021,Pesin2010,Imada1998,Tokura1993,Sachdev1999}. The convergence of these two fields presents a cutting-edge frontier for unraveling the study of interacting topological phases of quantum matter~\cite{Yankowitz2019,Kerelsky2019,Lisi2020,Cao2020,Serlin2020,Choi2021,Chen2021a,Tan2021,Jiang2021,Yang2020,Yu2021,Teng2022,Yin2022,Teng2023,Shao2023,Zhou2023}. However, a challenge in advancing this field is that electron-electron correlations are generally negligible in many well-known topological materials.  One effective strategy for enhancing electron-electron correlations involves introducing a high density of states in the vicinity of the Fermi level. Here the kagome lattice stands out as particularly promising candidate due to its unique electronic structures that features flat bands and van Hove singularities (VHSs). Although the initial focus was on two-dimensional (2D) lattices, recent work considers three-dimensional (3D) kagome systems in which the kagome layers coexist with other atomic layers in the unit cell~\cite{Ye2018,Liu2018,Yin2019,Ortiz2019,Yin2022a,Yang2023}. The focus has been a scenario in which the coupling along the stacking direction remains sufficiently weak so that the flat bands and VHSs of the kagome layers are preserved. Unfortunately, the complex interlayer couplings in these systems often lead to the disappearance of the distinctive kagome band characteristics.

The high symmetry of the kagome lattice also limits its potential applications. Recall that the breaking of symmetries has dramatically contributed to the emergence of numerous topological phases of quantum matter. For instance, chiral crystals that break inversion and mirror symmetries have been found to exhibit unique topological properties, including multi-fold chiral fermions and related quantized nonlinear optical responses~\cite{Juan2017,Chang2017,Tang2017,Chang2018,Sanchez2019,Rao2019,Takane2019,Yang2019,Schroeter2020}. Kagome KV$_3$Sb$_5$, with chiral charge order at low temperatures, has shown promise in realizing large orbital magnetization and unconventional superconductivity~\cite{Jiang2021,Luo2022,Li2022a}. However, prior to inducing strong correlation effects, KV$_3$Sb$_5$ exhibits non-chiral behavior at high temperatures. Novel correlated topological phenomena can thus be expected to arise from symmetry breaking in chiral crystals~\cite{Herbut2014,Jian2017,Li2022,Chiu2023,Sanchez2023,Rao2023}.



\begin{figure}[!htbp]
  \centering
  \includegraphics[scale=0.34,angle=0]{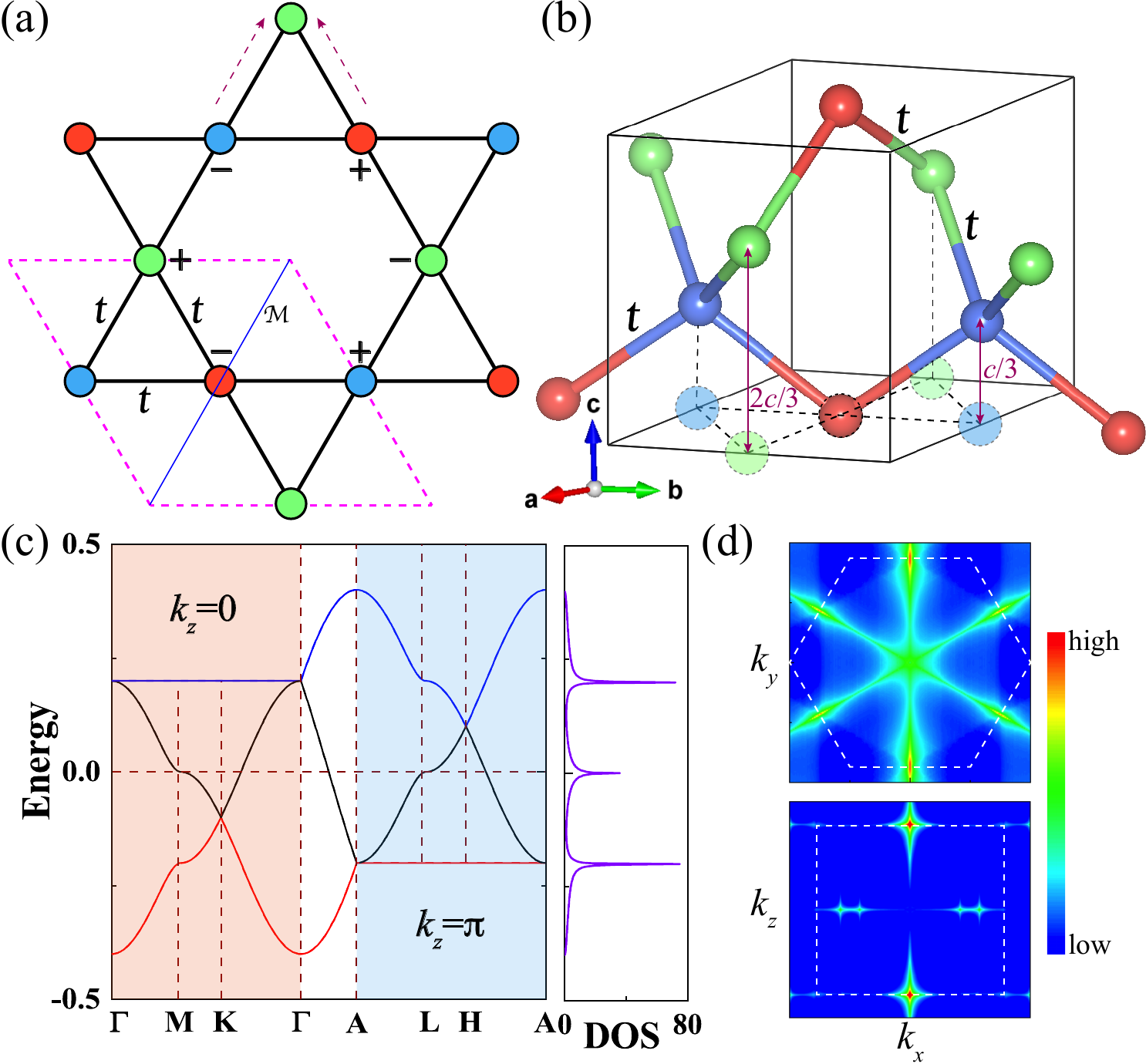}\\
  \caption{(a) A 2D kagome lattice with nearest-neighboring hopping $t$. Symbols '+' and '-' signify wave functions with equal amplitude but opposite phases, giving rise to compact localized states. Dashed arrows indicate canceled hoppings. (b) A 3D chiral kagome lattice, where its projection on the (001) plane is a 2D kagome. (c) Electronic band structure along the high-symmetry paths in the Brillouin zone, and the corresponding density of states (DOS). (d) $\mathbf{q}$-dependent static charge-density susceptibility $\chi_0(\mathbf{q})$ at low-temperature $T = 0.00001t$ with an enhanced amplitude at the CDW wave vectors $\mathbf{Q}$ on the (001) plane with ($k_1$, $k_2$, $k_3$) = (0, $\pi$, 0); ($\pi$, 0, 0); (-$\pi$, $\pi$, 0) (upper panel) and on the (010) plane with (0, 0, $\pi$) (bottom panel).}\label{fig1}
\end{figure}

The 2D kagome lattice comprises three atoms in the primitive cell represented by red, green, and blue spheres in Fig.~\ref{fig1}(a). The band structure of a single-orbital kagome model with only the nearest-neighboring (NN) hoppings exhibits a flat band, VHSs, and Dirac degeneracies, where one branch of the dispersive Dirac bands intersects the flat band at the $\Gamma$ point~\cite{Yin2019}. This is attributed to the wave function localization that forms compact localized states within the hexagon due to destructive interference. To extend this model to 3D, we can maintain the NN hopping but translate two atoms in 2D kagome lattice along the $z$ axis (out-of-plane direction) by $c$/3 and 2$c$/3, respectively, as depicted in Fig.~\ref{fig1}(b). In this configuration, it can be inferred that the system should preserve the characteristic kagome band features.

For the 2D kagome lattice, the green and blue atoms can be linked through mirror symmetry ($\mathcal{M}$). However, this symmetry is naturally broken when translating green and blue atoms along the $z$ axis at varying distances. Furthermore, unlike the 2D kagome lattice with inversion symmetry, the inversion centers for the three atoms in the 3D kagome lattice are not at the same height, thus resulting in inversion-symmetry-breaking. The 3D kagome lattice lacks both mirror and center inversion symmetries, classifying it as a chiral lattice possessing the screw symmetry $S_{6/3}$ with the combination of a six/three-fold rotation $C_{6/3}$ and a one-third fractional translation along the $z$ axis.

We now undertake an analytical verification to ascertain whether the kagome band characteristics remain preserved within the new lattice by constructing the Hamiltonian including only the NN hopping:
\begin{small}
\begin{equation}\label{eq1}
H_k\!\!=\!-2t\!\!\left(\begin{array}{ccc}\!d/t &\! {\rm cos}(\frac{k_2}{2})e^{-\frac{ik_3}{3}} &\! {\rm cos}(\frac{k_1+k_2}{2})e^{\frac{ik_3}{3}} \\ \!{\rm cos}(\frac{k_2}{2})e^{\frac{ik_3}{3}} &\! d/t & \!{\rm cos}(\frac{k_1}{2})e^{-\frac{ik_3}{3}} \\ \!{\rm cos}(\frac{k_1+k_2}{2})e^{-\frac{ik_3}{3}} &\! {\rm cos}(\frac{k_1}{2})e^{\frac{ik_3}{3}} &\! d/t\end{array}\right),
\end{equation}
\end{small}
where $\mathbf{a_1}=a(1, 0, 0)$, $\mathbf{a_2}=a(-1/2, \sqrt{3}/2, 0)$, $\mathbf{a_3}=c(0, 0, 1)$, and $k_i=\mathbf{k}\cdot \mathbf{a_i}$. Figure~\ref{fig1}(c) presents the electronic band structure of the new lattice described by Eq. (1) with the parameters $d=0$ and $t=0.1$. Indeed, as expected, the kagome bands are faithfully reproduced, including the flat band, VHSs, and the band degeneracies at $k_z (k_3)=0$ and $\pi$.  We also note that the energy bands in $k_z$= 0 and $\pi$ planes are reversed by a $\pi$ phase difference accumulated by the $z$-direction translation. Along $k_z$=0 to $\pi$, ladder-shaped bands connect the two kagome sets. We further calculate the density of states (DOS) for the system, similar to the 2D kagome system, revealing very sharp divergent peaks at the flat band and VHSs. This indicates potential strong electron-electron correlations at the energy of VHSs and flat bands. 

At the energy of VHSs, we compute the Lindhard function on the (001) and (010) planes, where exceptionally high-density electronic susceptibilities nested by these VHSs are observed [Fig.~\ref{fig1}(d)]. Within the (001) plane, the VHS at $M (L)$ favors particle-hole scatterings linked by the nesting vectors $\mathbf{Q}$, characterized by (0, $\pi$, 0), ($\pi$, 0, 0) and (-$\pi$, $\pi$, 0). Akin to 2D kagome materials, the enhanced susceptibility in the chiral kagome lattice can also prompt the emergence of an in-plane correlated charge density wave (CDW). In addition, because of the existence of a third dimension, on the (010) plane, the nearest VHSs at $M$ and $L$ in the $k_z$ direction can be further associated with the vector $\mathbf{Q}$(0, 0, $\pi$), suggesting a potential CDW along the $z$ direction.

We note that our formulation is applicable not only to electronic structures but it can also handle phononic lattices, see Supplemental Materials (SM) and Fig. S1 for details. Our results suggest the presence of strong electron-electron and electron-phonon interactions in the chiral kagome lattice.
 
\begin{figure}[!htbp]
  \centering
  \includegraphics[scale=0.31,angle=0]{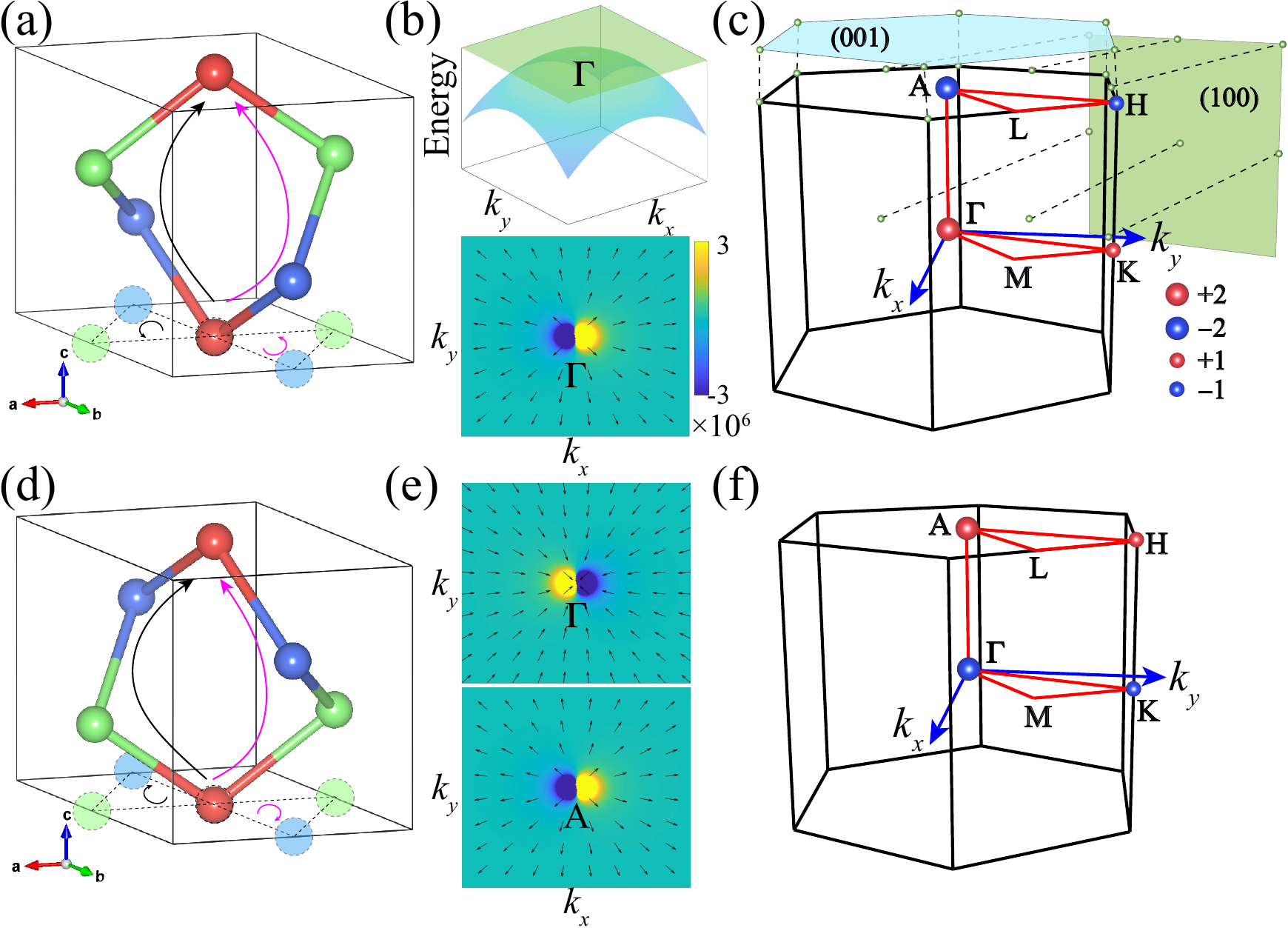}\\
  \caption{(a) Right-handed Crystal structure, (b) 3D energy dispersion of the degeneracy at the $\Gamma$ point, along with Berry curvature $\Omega_x$ around the $\Gamma$ points on the (001) plane with flow direction indicated, and (c) distribution of Weyl points in the Brillouin zone (BZ), where colored spheres with different sizes label Weyl points with different topological charges, and associated projected surface BZs for the 3D chiral kagome lattice with right-handedness. In (a), the upward arrow illustrates the evolution process of atoms moving from the plane along the $c$ direction, while the spiral arrow projected onto the bottom surface indicates chirality. (d) Right-handed Crystal structure, and its (e)  Berry curvature $\Omega_x$ around the $\Gamma$ and $A$ points, and (f) distribution of Weyl points for the 3D chiral kagome lattice with left-handedness.}\label{fig2}
\end{figure}

Based on the preceding considerations, we have constructed a 3D chiral kagome lattice. To discuss topological effects arising from symmetry breaking, we first consider a right-handed crystal [Fig.~\ref{fig2}(a)], where the atoms follow an anti-clockwise rotation with respect to the $z$ axis following the right-hand rule. By analyzing the Berry curvature of the bands, we observe that the two-fold degeneracies at the $\Gamma$, $A$, $K$, and $H$ points possess quantized Chern numbers. Specifically, in the right-handed kagome lattice, a Chern number $C=2$ is ascribed to the $\Gamma$ point, serving as a source of Berry curvature [Fig.~\ref{fig2}(b)]. Each of the two equivalent $H$ points carries a Chern number $C=-1$ to compensate for the positive chiral charge at the $\Gamma$ point [Fig.~\ref{fig2}(c)]. Similarly, the right-handed 3D kagome lattice exhibits a $C=-2$ at the $A$ point and $C=1$ at the $K$ point [Fig.~\ref{fig2}(c)].  Furthermore, we observe that one of the branches of the double Weyl cone at both $\Gamma$ and $A$ points comprises flat bands [Fig.~\ref{fig2}(b), top panel]. 

Applying a vertical mirror operation to the right-handed 3D kagome lattice [Fig.~\ref{fig2}(a)] results in the lattice with left-handedness  [Fig.~\ref{fig2}(d)], wherein the atoms rotate clockwise along the $z$ axis. Interestingly, the electronic dispersions of the left-handed crystal exactly mirror those of the right-handed counterpart, with both structures representing ground states. However, a significant distinction arises in the flow of Berry curvature in momentum space, which undergoes a complete reversal due to the change in structural chirality in real space [Figs.~\ref{fig2}(b),(e)]. This reversal, in turn, affects the sign of the Chern numbers associated with the massless Weyl fermions [Figs.~\ref{fig2}(c),(f)]. Our analysis demonstrates that the topological Chern number in the kagome lattice is inherently dictated by its structural chirality. Consequently, by tuning the handedness of the lattice, one gains the ability to manipulate the Chern numbers and thus the related topological quantum responses, including the direction of photocurrents ~\cite{Juan2017}. 

\begin{figure}[!htbp]
  \centering
  \includegraphics[scale=0.35,angle=0]{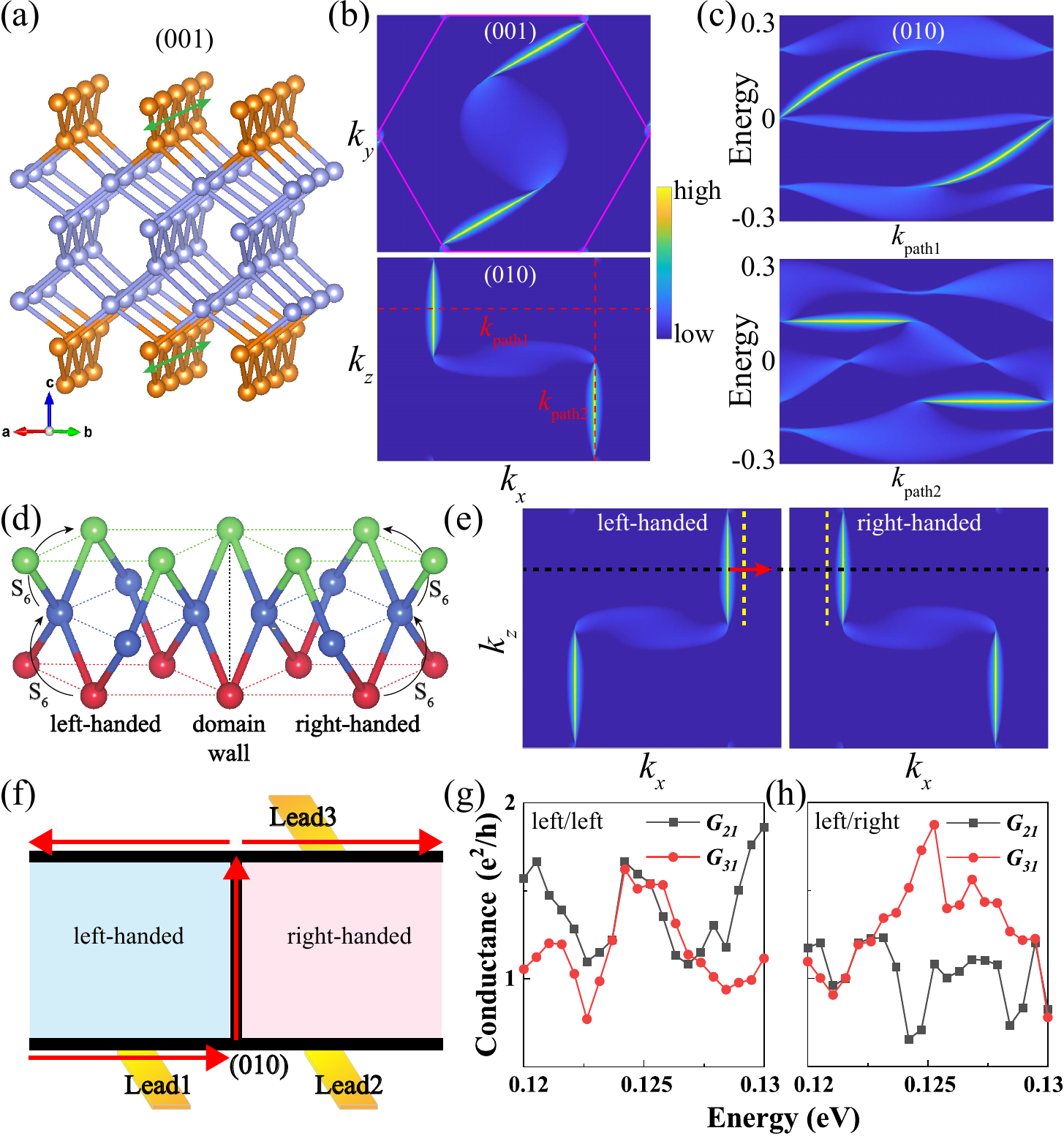}\\
  \caption{(a) Schematic diagram of 1D channels on the (001) surface, where the green arrows represent the 1D channels. (b) The (001) and (010) surface state spectral functions at 0.12 eV, where a path perpendicular to Fermi arcs ($k_{\rm path1}$) and a path parallel to Fermi arcs ($k_{\rm path2}$) are highlighted in (b). (c) The surface spectral functions along the specific $k$ paths for the (010) surface. (d) The domain wall formed by different-handed structures that are mirror symmetrical about the domain wall. (e) The current transport at the interface (domain wall) on the (010) plane. Solid curves are fermi arcs at 0.12 eV, while dashed curves are fermi arcs at slightly higher energy. (f) Schematic diagram of non-local transport for finite size materials and three-terminal device, where the red arrows indicate the propagation process of surface current. Lead1 and Lead2 are connected to the (010) plane, and Lead3 is connected to the (0$\bar{1}$0) plane. The red and blue regions represent chiral opposite lattices. (g) and (h) The conductance of devices with uniform chirality and opposite chirality (corresponding to (f)) in the energy range of (010) surface states (around 0.125 eV), respectively. $G_{2(3)1}$ is the conductance between Lead1 and Lead2(3).}\label{fig3}
\end{figure}

We now turn to discuss the unique topological surface responses of our 3D chiral kagome lattice. Typically, the surfaces of a 3D material exhibit 2D electronic structures. However, intriguingly, despite the inherent 3D nature of our chiral kagome lattice, its surfaces manifest distinct one-dimensional (1D) features. This is evident on the (001) surface of the lattice, wherein conducting electrons from the top layer are bound to the inner layer through NN hopping along a 1D chain [Fig.~\ref{fig3}(a)]. Consequently, the emergence of 1D Fermi arc surface states is anticipated. Indeed, on the constant energy contours of the (001) surface, we observe the presence of two 1D Fermi arc surface states [Fig.~\ref{fig3}(b), top panel]. This distinctive 1D nature of Fermi arcs also extends to other terminations, such as the (010) surface, signifying the general occurrence of 1D surface states in the 3D chiral kagome lattice [Fig.~\ref{fig3}(b), bottom panel]. A detailed analysis of the energy dispersion of the Fermi arc surface states reveals that they exhibit dispersion in only one direction [Fig.~\ref{fig3}(c), top panel], while remaining completely flat in the perpendicular direction [Fig.~\ref{fig3}(c), bottom panel]. The surface states on the (001) plane also display dispersion in one direction while remaining flat in the orthogonal direction [Fig. S2]. This further substantiates the 1D nature of these topological Fermi arcs in our 3D chiral kagome lattice. 

The presence of 1D surface states characterized by parallel Fermi arcs suggests a pronounced Fermi nesting effect, hinting at potential strong correlation effects on the surface of the chiral kagome lattice. It is essential to recognize that while both the bulk and surface states can exhibit strong correlations, the underlying mechanisms driving these effects are fundamentally distinct. The correlation of bulk states originates from VHSs and flat bands, whereas for surface states, it arises from the specific 1D configurations on the surface. As a consequence, it is plausible that two independent correlation effects are induced, operating separately in the bulk and on the surface of the chiral kagome lattice. For instance, considering the CDW phenomenon as an illustration, the VHSs in bulk may lead to a 2$\times$2$\times$1 or 2$\times$2$\times$2 CDW order [Fig.~\ref{fig1}(d)] ~\cite{Jiang2021,Teng2022}. In contrast, on the surface, the $\mathbf{Q}$-vector characterizing the charge order could acquire a relatively arbitrary value, which is determined by the separation of Fermi arcs \cite{Li2022,Sanchez2023,Rao2023}. The distinct behaviors of bulk and surface correlations open new possibilities for correlated topological effects on the chiral kagome lattice.


In addition to the potential strong correlations induced on the surface, the presence of 1D Fermi arcs in the 3D chiral kagome lattice offers opportunities for realizing exotic nonlocal transport phenomena. Unlike its 2D counterpart with inversion symmetry, the 3D chiral kagome lattice can readily form left- and right-handed chiral domains that are mirror symmetric on either side of the domain wall [Fig.~\ref{fig3}(d)]. 
The connectivity of the Fermi arcs is also reversed on the two sides of the domain wall, as shown by the (010) surface states in Fig.~\ref{fig3}(e). Consider the following example, when a current originates from the Fermi arc of the left domain and moves toward the right domain [Fig.~\ref{fig3}(e)], it encounters a unique situation. Due to the inherent absence of reflection surface modes, this current cannot be reflected back. Additionally, it cannot pass through the domain wall to reach the (010) surface of the right domain, as the group velocity of states on the right domain opposes the electron's direction of motion. Consequently, the current can only be transmitted along the interface of the domain wall, entering the right domain from the surface on the other side [Fig.~\ref{fig3}(f)]. The current transport on the (001) surface can exhibit negative reflection and refraction as discussed in Fig. S3.

In conventional diffusion electronics, transport adheres to Ohm's law, where the conductance decreases with increasing separation between two leads. However, in our proposed scenario, we expect the conductance to offer novel perspectives for designing materials with tailored transport properties and potential applications in advanced electronic devices. Surprisingly, we anticipate the conductance between Lead2 and Lead1 to be smaller than the conductance between Lead3 and Lead1, even though Leads1 and 3 are more widely separated from each other [Fig.~\ref{fig3}(f)]. To verify this picture, we directly calculate the conductance of two devices of the same size for comparison [Figs.~\ref{fig3}(g),(h)], one device has a uniform chirality, while the other consists of two regions with opposite chirality [Fig.~\ref{fig3}(d)]. In the 3D kagome device with uniform chirality, the conductance $G_{21}$ is larger than $G_{31}$ [Fig.~\ref{fig3}(g)], indicating that the incoming current from Lead1 tends to partition more towards Lead2. This preference can be attributed to Lead2 being closer to Lead1 compared to Lead3. In contrast, in the device consisting of two chiralities, the conductance $G_{31}$ is larger than $G_{21}$ [Fig.~\ref{fig3}(h)]. Our calculations indeed reveal the emergence of unconventional nonlocal transport effects in the 3D chiral kagome lattice because of the lattice's chirality and unique 1D Fermi arcs. These findings open up new avenues for engineering materials with intriguing transport properties, suggesting the potential for exciting applications in future electronic devices.


The most basic symmetry operation in the 3D chiral kagome lattice is the screw symmetry $S_6$ (or $S_3$). Space groups (SGs) that satisfy this symmetry include SGs 144, 145, 151-154, 171, 172, 180 and 181. By consulting the Inorganic Crystal Structure Database (ICSD)~\cite{Levin2020}, we discover a large number of experimentally synthesized materials that hold the 3D chiral kagome lattice with the 3D kagome band characteristics [Fig.~\ref{fig1}(c)] in their electronic or phonon band structures, such as CePO$_4$, TiC$_4$H$_8$NO$_{10}$, BeF$_2$, BaCoO$_2$, BaZnO$_2$, GeO$_2$ and CuH$_{20}$N$_6$OF$_2$, InH$_8$C$_4$NO$_{10}$, SiO$_2$ and KScH$_4$(C$_2$O$_5$)$_2$, and Li(BH)$_5$ as shown in Figs. S4 and S5. Among these known materials, we find various types, including metals, insulators, nonmagnets, and ferromagnets. This diversity highlights the potential of 3D chiral kagome materials for a wide range of applications and investigations across different material types and properties.

\begin{figure}[!!htb]
  \centering
  \includegraphics[scale=0.54,angle=0]{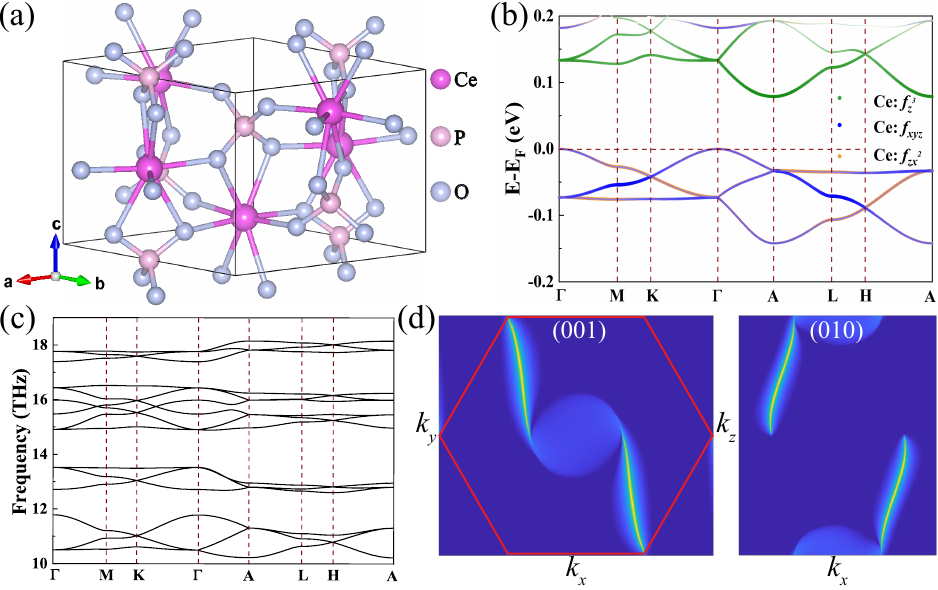}\\
  \caption{(a) Crystal structure of CePO$_4$ in SG 180, where Ce atoms form a 3D chiral kagome lattice. (b) Electronic band structure and (c) phonon spectrum of CePO$_4$. (d) The (001) and (100) surface state spectral functions at -0.04 eV.}\label{fig4}
\end{figure}

Finally, we focus on a representative material candidate for the 3D chiral kagome lattice CePO$_4$, where the 3D chiral kagome lattice is composed of Ce atoms [Fig.~\ref{fig4}(a)]. This material is a ferromagnetic semiconductor with a magnetic moment of about 1 $\mu_B$ per Ce atom. Interestingly, only one spin species (spin up) is distributed near the Fermi level, while the other one remains far away from the Fermi level [Fig.~\ref{fig4}(b)]. The three valence bands near the Fermi level in CePO$_4$ form the ideal 3D kagome bands, characterized by flat bands with a bandwidth of less than 3 meV on $k_z=0$ and $\pi$ planes. By introducing about 0.8 holes through doping, the Fermi level can be effectively tuned to align with the flat band. This can be achieved, for instance, by substituting part of the P atoms in CePO$_4$ with Si or C atoms. The phonon spectrum of CePO$_4$ further adds to its unique properties, revealing several groups of 3D chiral kagome bands in the frequency range of 10 to 18 THz [Fig.~\ref{fig4}(c)]. It is worth noting that although our model only considers the NN hopping, the robust nature of the flat bands for electrons and phonons in real materials suggests that our model is highly compatible with the behavior exhibited by the real materials. Additionally, the surface states of CePO$_4$ exhibit 1D Fermi arcs [Fig.~\ref{fig4}(d)], which is consistent with our model. This distinctive combination of electronic and phonon properties in CePO$_4$ positions it as a promising candidate for exploring novel quantum phenomena and potential applications in the realm of 3D chiral kagome materials.

\section*{Acknowledgement}
Work at Nanyang Technological University was supported by the National Research Foundation, Singapore under its Fellowship Award (NRF-NRFF13-2021-0010) and the Nanyang Assistant Professorship grant (NTUSUG). Work at the National University of Singapore was supported by the Ministry of Education, Singapore, under its MOE AcRF Tier 3 Award MOE2018-T3-1-002. The work at Northeastern University was supported by the Air Force Office of Scientific Research under award number FA9550-20-1-0322 and benefited from the computational resources of Northeastern University's Advanced Scientific Computation Center (ASCC) and the Discovery Cluster. The work at TIFR Mumbai was supported by the Department of Atomic Energy of the Government of India under Project No. 12-R$\&$D-TFR-5.10-0100 and benefited from the computational resources of TIFR Mumbai.

J-Y. Y., X. Z., and T. H. contributed equally to this work.

%

\end{document}